# Higher dispersion and efficiency Bragg gratings for optical spectroscopy

Will Saunders[a,b,1], Kai Zhang[c], Thomas Flügel-Paul[d]

[a] Australian Astronomical Observatory, North Ryde, NSW, Australia,
[b] Australian Astronomical Optics, Macquarie University, NSW, Australia
[c] National Astronomical Observatories, Nanjing Institute of Astronomical Optics & Technology, Chinese Academy of Sciences, Nanjing, China;
[d] Fraunhofer Institute for Applied Optics and Precision Engineering IOF, Jena, Germany

## ABSTRACT

Massively multiplexed spectroscopic stellar surveys such as MSE present enormous challenges in the spectrograph design. The combination of high multiplex, large telescope aperture, high resolution (R~40,000) and natural seeing implies that multiple spectrographs with large beam sizes, large grating angles, and fast camera speeds are required, with high cost and risk. An attractive option to reduce the beam size is to use Bragg-type gratings at much higher angles than hitherto considered. As well as reducing the spectrograph size and cost, this also allows the possibility of very high efficiency due to a close match of *s* and *p*-polarization Bragg efficiency peaks. The grating itself could be a VPH grating, but Surface Relief (SR) gratings offer an increasingly attractive alternative, with higher maximum line density and better bandwidth. In either case, the grating needs to be immersed within large prisms to get the light to and from the grating at the required angles. We present grating designs and nominal spectrograph designs showing the efficiency gains and size reductions such gratings might allow for the MSE high resolution spectrograph.

## 1. INTRODUCTION

The MSE project [1] is a proposal to replace the CFHT with a dedicated 10m-class wide-field spectroscopic telescope. The telescope would feed 4300 fibers, of which one third are for high resolution (HR) use. The HR fibers feed two dedicated spectrographs, each with three arms covering Blue (401-417nm), Green (471-489nm) and Red (625-674nm) regions of the spectrum. Simple considerations of telescope aperture, proposed fiber aperture size (~0.8″) and required resolution (40K for Blue and Green arms, 20K for Red) indicate that very large beam-sizes and gratings are required, together with large grating angles and fast cameras. Transmission gratings are strongly preferred by the efficiency and field angle requirements, and because there are to be 500+ spectra on each detector. Currently, the largest optical VPH gratings for astronomical use are for the HERMES spectrograph on the AAT [2], 200mm × 500mm, while larger gratings have been made for NIR use in the APOGEE spectrographs [3], 300mm × 500mm. Both sets of gratings were made by Kaiser Optical Systems, Inc. (KOSI). For MSE, the AΩ of the beam emerging from the fibers is almost identical to that for HERMES (11m × 0.8″ vs 3.9m × 2.2″), but the required resolution is 43% larger. This increase must come from some combination of a larger beam size, a larger grating angle, or immersing the grating. Immersing the grating between two large right-angled prisms seems by far the least difficult way to achieve the required resolution. It allows much higher Bragg angles (the grating angle within the grating itself) while avoiding total internal reflection (TIR). However, immersed gratings require a higher line density.

At large grating angles and line densities, the optimum parameters for the grating structure (refractive index, refractive index modulation, thickness) differ for *s* and *p*-polarizations. This means that in general, the grating can offer good efficiency in only one polarization (and this is true for HERMES). MSE's science requirements demand excellent overall efficiency, especially in the Blue arm, so this is not an option. However, the grating efficiency is periodic with thickness, and this allows excellent peak efficiency to be offered simultaneously in both polarizations, by matching an efficiency peak in one polarization with a *different* peak in the other. This restricts the Bragg angle (the grating angle within the grating structure itself) to certain discrete values, as discussed below. These '*s-p* phased' [4] or 'High Throughput' [5] or 'Dickson' [6] gratings have been made since (at least) 2003, and a patent has been registered covering some of them [7].

---

[1] *will.saunders@aao.gov.au*

The current design for the MSE HR spectrograph is presented in detail by Zhang et al [8]. It consists of 3 arms with 300mm beam size and f/1.5 cameras with 91mm × 91mm detectors. The VPH gratings are used at Bragg angles ~60° (Blue/Green) and ~41° (Red), meaning they cannot be very well phase-matched. The grating size is determined by the capabilities of KOSI, but still has some vignetting. The overall spectrograph size is 3m x 3m x 0.8m, the largest lenses are 500mm aperture (restricting the glass and polisher choices), and the glass mass is ~400kg. Size and weight are particular concerns for MSE, because both are very limited on the Nasmyth platforms. If the spectrographs cannot be accommodated there, they must go below the telescope floor with an azimuthal fiber wrap, a significant increase in fiber length and consequent loss of throughput, especially at the blue wavelengths where it is most crucial.

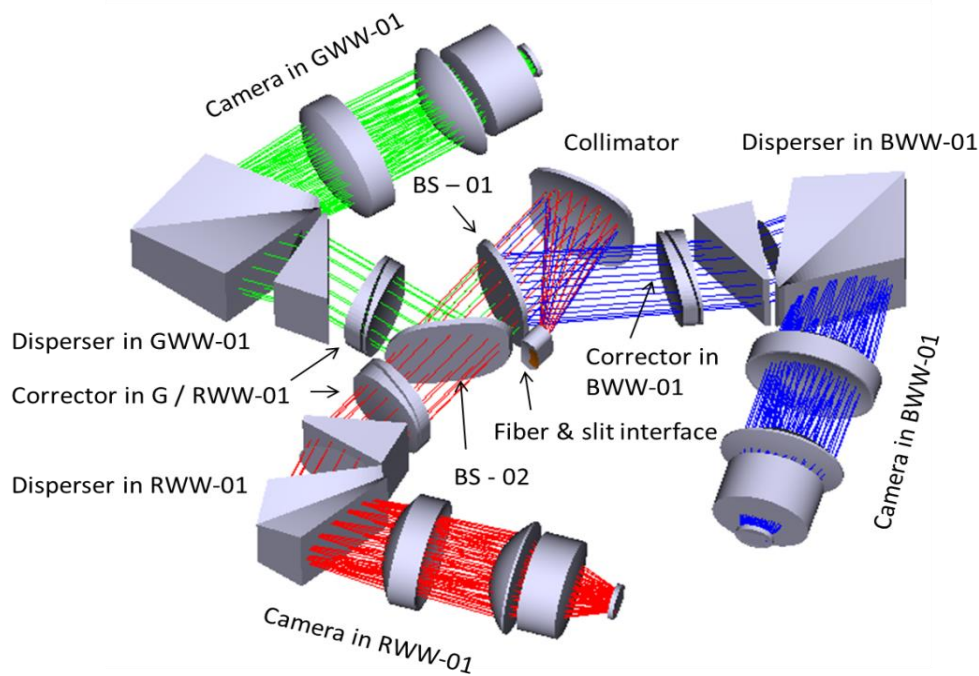

**Figure 1. Current MSE HR spectrograph design, from Zhang et al 2018 [8].**

This design is already an evolution of the CoDR designs, which were deemed to have significant technical risks. It remains very challenging, and fails to meet the very demanding Blue throughput requirements. Therefore, it would be very attractive to find a design with higher grating angles and smaller beam size, which could simultaneously reduce size, weight and cost while increasing efficiency via the used of *s-p* phased gratings.

A new development is the rapid evolution of surface relief (SR) gratings [9,10]. In principle, these offer higher line densities and refractive index variations (and hence better bandwidths) than VPH gratings can offer, and so may in time supplant them as the dispersers of choice.

## 2. BEAM SIZE

Spectrograph cost/difficulty/risk strongly driven by the collimated beam size. For an immersed grating with the chief ray at right-angles to the prism input and exit faces (Figure 2),

$$B = R\,D_T\,\phi_F / (2n_1 \tan\alpha_0) \qquad (1)$$

where $R$ is the resolution, $D_T$ is the telescope diameter, $\phi_F$ is the FWHM angular slit width on the sky, $n_1$ is the index of the immersion medium, and $\alpha_0$ is the overall grating angle. For an unimmersed grating, $n_1 = 1$, showing showing immediately that immersing the grating greatly reduces the beam size for fixed grating angle.

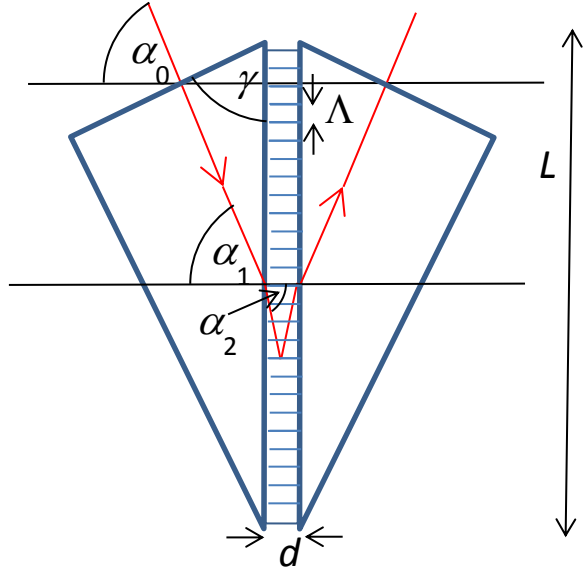

**Figure 2. Schematic overall grating layout referred to throughout this paper. The grating structure (whether VPH or Surface Relief) is assumed to be unslanted, and the prisms are assumed to have input faces orthogonal to the beam.**

From equation (1), $B$ scales directly with resolution, telescope size, and fiber aperture, and inversely with immersion medium index. All these are fixed (or at least strongly constrained) by other requirements, leaving only the Bragg angle as a relatively free parameter. Hence, to reduce the beam size, a larger grating angle is required. But note that the *length* of the grating is given by $L = B/\cos\alpha_1 = R\,D_T\,\phi_F / (2n_1 \sin\alpha_0)$, so there is an irreducible minimum grating length $\sim R\,D_T\,\phi_F/3$. For MSE, this is ~585mm, already larger than the 500mm length of the HERMES or APOGEE gratings.

But for MSE, efficiency also paramount, especially in the Blue arm. The combined requirement for high efficiency and high grating angle pushes us very strongly towards an *s-p* phased grating.

### 3. 'SUPERDICKSON' GRATINGS

The peak efficiency of VPH gratings was approximated by Kolgenik [11]. Following the notation of Baldry et al [4],

$$\eta \approx \tfrac{1}{2}\sin^2\!\left(\frac{\pi\,\Delta n_2\,d}{\lambda\cos\alpha_2}\right) + \tfrac{1}{2}\sin^2\!\left(\frac{\pi\,\Delta n_2\,d}{\lambda\cos\alpha_2}\cos 2\alpha_2\right) \qquad (2)$$

where $\lambda$ is the wavelength, $\Delta n_2$ is the index modulation, $d$ is the dichromated gelatin (DCG) thickness, $\alpha_2$ is the grating angle *within the DCG*, and the two terms are for *s* and *p* polarizations respectively.

For either polarization, we can get ~100% efficiency by suitable choice of $d$ and $\Delta n_2$, to get $\pi/2$ (or $3\pi/2$, $5\pi/2$ etc) within the brackets. For small angles, $\cos 2\alpha_2 \sim 1$, and hence excellent peak efficiency is possible in both polarizations simultaneously. But as $\alpha_2$ increases, the $\cos(2\alpha_2)$ term introduces a mismatch between the desired DCG properties for each polarization. Simultaneous high efficiency for both polarizations is still possible for special values of $\alpha_2$, by matching an efficiency peak in the *s* polarization with a *different* peak in the *p* polarization. These are '*s-p* phased' or 'High Throughput' or 'Dickson' gratings.

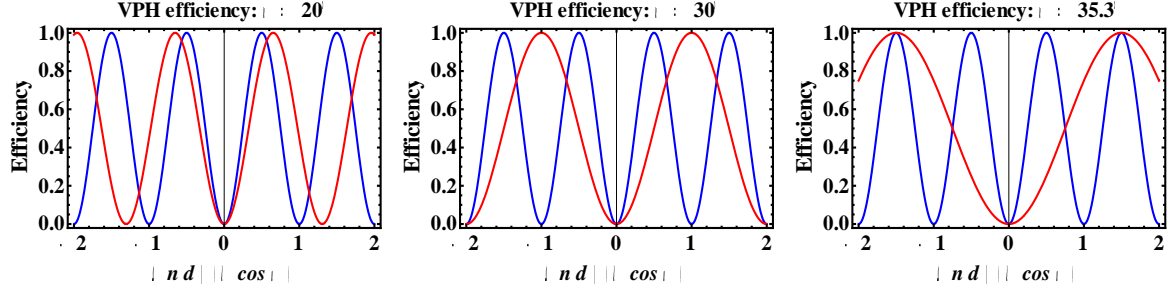

**Figure 2. Peak efficiency vs thickness for *s* and *p* polarizations (shown in blue and red respectively), for Bragg angles 20°, 30°, 35.3°. As the Bragg angle increases, there is an increasing mismatch between the peaks, but at 35.3°, excellent efficiency in both polarizations can be obtained simultaneously by matching the 1st *p*-peak with the 2nd *s*-peak, i.e. making a grating with $\Delta n_2\, d = 3\,\lambda\,\cos 35.3°/2 = 1.224\,\lambda$.**

For this to happen, we need

$$\frac{2\,\Delta n_2\, d}{\lambda \cos\alpha_2} = 2a+1 \qquad (3)$$

and

$$\cos 2\alpha_2 = \frac{2b+1}{2a+1} \qquad (4)$$

for integral *a*, *b*.

The first such grating is obtained by matching the 1st *p*-peak with the 2nd *s*-peak, i.e. $(a,b) = (0,1)$. This happens when $\cos 2\alpha_2 = 1/3$, or $\alpha_2 = 35.3°$ (Figure 2). As far as we are aware, the first astronomical uses of such gratings were for the 6dF/RAVE project on the UK Schmidt (2003) [12] and AAOmega on the AAT (2004)[13], both for CaII triplet work (~850nm).

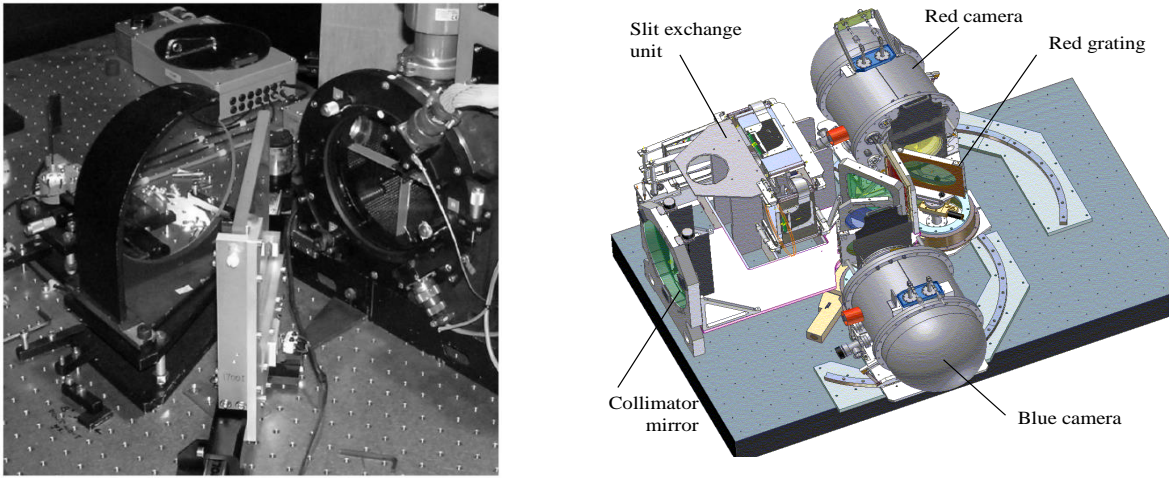

**Figure 3. 6dF/RAVE and AAOmega spectrographs in high resolution mode, with Dickson gratings made by Richard Rallison**

Wasatch took a patent on Dickson gratings in 2004 [7], specifically covering non-negative integral *a* and *b*. It's obvious that *a* must be non -ve, from equation (3). However, *b* is not so constrained, **and there are multiple families of further solutions with –ve *b*.** The most interesting solutions are summarized in Table 1.

**Table 1. Properties of *s-p* phased VPH gratings. Only the obviously most interesting are included, any positive integral *a* and integral *b* gives an *s-p* phased grating.**

| *a* | 0 | 1 | 2 | 1 | 2 | 0 |
|---|---|---|---|---|---|---|
| *b* | ~0 | 0 | 0 | -1 | -2 | ~-1 |
| $(b+\frac{1}{2})/(a+\frac{1}{2})$ = $cos(2\alpha_2)$ | ~1 | 1/3 | 1/5 | -1/3 | -3/5 | ~-1 |
| $\alpha_2$ | ~0 | 35.3° | 39.2° | 54.7° | 63.4° | ~90° |
| $tan(\alpha_2)$ | small | $1/\sqrt{2}$ | $\sqrt{(2/3)}$ | $\sqrt{2}$ | 2 | large |
| Notes | Normal VPH gratings | Classic Dickson | Higher-order Dickson, narrow bandwidth | New design, twice Dickson resolution | High resolution but narrow bandwidth | Arbitrary dispersion, limited only by TIR |

All the new (-ve *b*) solutions have Bragg angles > 45°. This means they all need prisms to get the light into and out of the grating while avoiding Total Internal Reflection (TIR). That is, these are necessarily immersed gratings. This also increases the resolution (for fixed angle between input and output beams), and reduces air/glass surface losses at input and output, so seems like a very desirable feature, even though these prisms will be heavy (many tens of kg for typical beam sizes and angles). It is assumed that the prisms have input faces orthogonal to the incoming/outgoing beams, since this gives the highest possible resolution without increasing the required pupil relief.

Kogelnik also gave an approximate formula for the FWHM bandwidth,

$$\frac{\Delta\lambda}{\lambda} \sim \frac{\Lambda}{d \tan \alpha_2} \qquad (5)$$

where $\Lambda$ is the grating period (so the bandwidth is the inverse of the number of fringes seen by $0^{th}$ order rays). The formula is not quantitatively usefully at large grating angles, but shows that for the best bandwidth, we want the smallest possible *d*. Since $\Delta n_2 d$ is constrained from equation (2), this means we want in general the largest possible $\Delta n_2$ that the technology allows. Equation (3) also shows that the smallest value of $\Delta n_2 d$ occurs when *a* = 0. For any other value of *a* (and given that the maximum $\Delta n_2$ is fixed), this means that the grating thickness is increased, and hence the bandwidth is in general decreased for *s-p* phased gratings. For *a* = 0, equation (4) gives no exact solutions. But for *b* = 0, equation (4) is close to being satisfied for small $\alpha_2$. This is why normal unphased VPH gratings work so well at moderate grating angles. But also, for *b* = -1, equation (4) is equally close to being satisfied when $\alpha_2$ is close to 90°. That is, *VPH gratings can in principal work as well at high dispersion as they do at low dispersion.*

The two most interesting new classes of gratings are **(*a*, *b*) = (1,-1)** and **(*a*, *b*) ~ (0,-1)**, are these are discussed below.

**(*a*,*b*) = (1,-1), $\alpha_2$ = 54.7°**

This solution is mentioned by Baldry et al, and has been made in small size by KOSI for dense wavelength multiplexing [14]. It gives a grating with twice the resolution of a classic Dickson grating. It appears to be manufacturable as a VPH grating for optical astronomical use, though the preferred DCG thickness is somewhat thinner than current gratings. The theoretical efficiency for a nominal design is shown in Figure 4. KOSI has provided a preliminary efficiency curve for a similar grating, with comparable performance. For MSE use (401nm-417nm), the efficiency profile is reasonable (>65% everywhere), but to get the required dispersion requires a beam size larger than KOSI's current limit of 304mm.

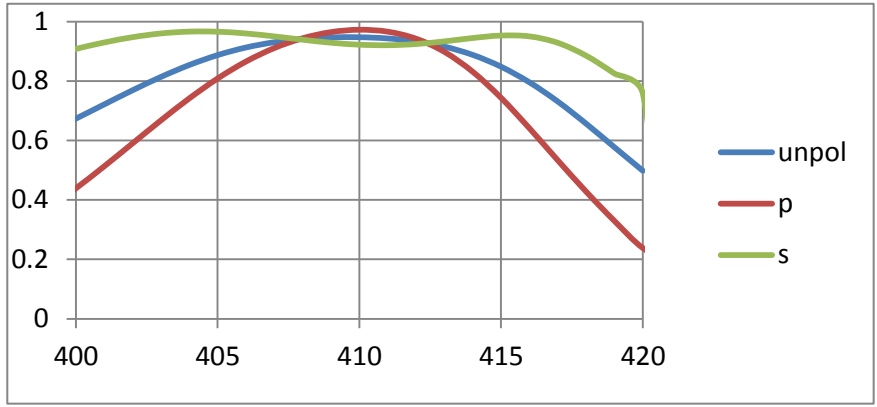

**Figure 4. Theoretical *s*,*p* and unpolarized efficiencies for phase-matched 5378/mm grating with Bragg angle ~54.7°.**

**(*a*,*b*) ~ (0,-1), $\alpha_2$~90°**

This solution is unphysical in its exact form, since it would imply a Bragg angle of 90° (so no transmission). But the situation is analogous to that at low dispersion (where excellent performance is obtained for Bragg angles up to ~25° or so), and excellent performance can be obtained when the Bragg angle is ~65° or greater, and the closer to 90°, the better the efficiency. Thus this solution potentially offers superb efficiency and unlimited resolution. The issues are the practical ones of achieving the required line densities (>6000/mm for a grating working around 400nm), achieving the very thin DCG layers that are wanted, and getting the light into and out of the grating without TIR.

For VPH gratings recorded with a laser wavelength of 488nm on fused silica substrates, there is a hard line density limit of ~6050/mm, just from the grating equation. Higher density substrates allow higher densities, but cause light losses at blue wavelengths. One solution would be a bluer laser wavelength, but this is difficult to work with. Figure 5 shows the potential performance of a 6445/mm VPH grating, if this line density could ever be achieved. The resolution is now adequate for MSE, but both peak efficiency and bandwidth are less good than the design shown in Figure 4. Efficiency is poor at the blue end, just where it is most critical.

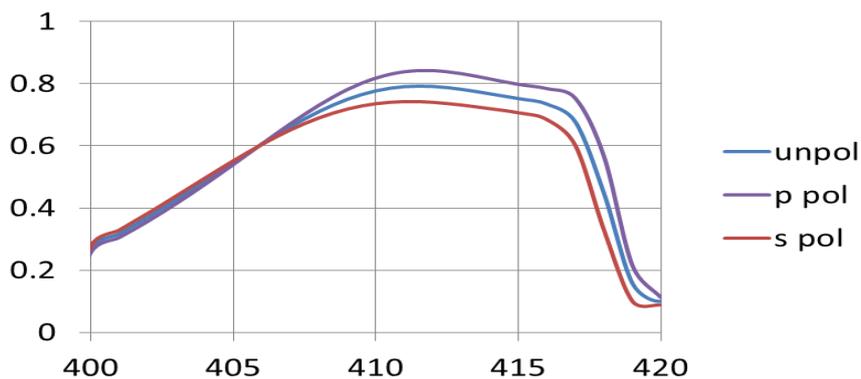

**Figure 5. Theoretical *s*,*p* and unpolarized efficiencies for a nominal 6445/mm VPH grating with Bragg angle ~72°.**

KOSI have offered a speculative solution at 6100/mm with better bandwidth, consisting of parameters (thickness, line density, size, index variation) that have all been achieved separately, but not together. Achieving this line density would require BK7 (or similar) substrates and prisms, incurring a modest (5-10%) absorbtion loss.

**Mosaicing**

Either grating design would require mosaicking, in at least the spectral direction. A parallelism of a few arcsec is required, making multiple exposures on the same substrate risky. If multiple substrates are used, they would need to be of very accurately matched thicknesses.

To summarize: it is not clear that VPH gratings offer a disperser solution for MSE. In any case, extensive prototyping and development would be required, both for the grating structure and for mosiacing into the required sizes.

## 4. SURFACE RELIEF GRATINGS

SR gratings [9,10] offer a very promising alternative disperser technology. A series of grooves with rectangular cross-section are recorded in a fused silica substrate by lithographic methods. Both the grooves and the walls between them can have large aspect ratios (ie depth to width ratio). The grooves can be filled in (to make a planarized grating) to increase both the refractive index and the refractive index contrast. When capped with a fused silica superstrate, the resulting grating is as robust as a VPH grating. Recording speeds have recently increased rapidly, making large area astronomical gratings feasible. Current size limits for Fraunhofer IOF are 270mm x 130mm rectangular, so mosiacing is necessary in the spectral direction (with the same few arcsec precision rquirement as for VPH gratings), and beam-size is somewhat constrained. However, compared with VPH gratings, SR gratings offer some significant differences, mostly positive:

- Higher line densities can be achieved than for VPH gratings. Current technology at Fraunhofer IOF allows ~6500/mm; the limiting factor being the minimum groove width of ~108nm combined with the requirement for a reasonable aspect ratio (<~8) for the walls.

- If the grating is planarized with a high refractive index material such as $TiO_2$, then very large refractive index variations are possible. This means the gratings can be very thin (less than one wavelength), giving exceptional bandwidth.

- The average refractive index within the grating structure is also very high. This means (just from Snell's law) that the Bragg angle is much reduced, for fixed dispersion. In general, this is advantageous.

- To avoid excessive Fresnell losses between the substrate and the grating, a thin film of some material with intermediate refractive index (such as $Al_2O_3$) can be added to one or both sides of the grating structure.

- The Kogelnik approximation is not useful for these gratings, because the refractive index variation is so large. Efficiencies must be modelled by full RCWA. This means the phasing arguments of Section 3 are also not relevant. However, the bandwidths are so good, that achieving excellant *s* and *p* efficiency simultaneously is very much easier.

The schematic layout of such a surface relief grating is shown in Figure 6.

Fraunhofer IOF has provided theoretical performances for two gratings for use at ~410nm, at 5700/mm and 6450/mm. Both are etched in fused silica, with the rulings planarized with $TiO_2$. Both offer potentially superb performance, essentially because the index variations are so large, and this determines the bandwidth. Equally good performance is achieved at the lower resolution required for the Red arm.

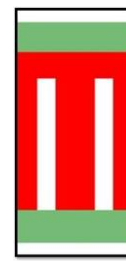

**Figure 6. Proposed schematic surface relief grating structure. White is fused silica, red is $TiO_2$, green is $Al_2O_3$.**

Extensive development would be required to achieve the required tolerances, both for the gratings structure itself and for mosiacing. However, the implications for the overall survey efficiency are so compelling, that such development is intended, as soon as possible.

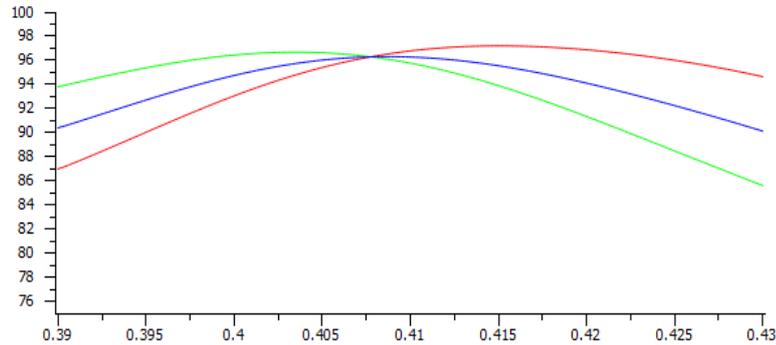

**Figure 7. Theoretical efficiency of a proposed 5700/mm planarized surface relief grating from Fraunhofer IOF. The efficiency for unpolarized light is shown in blue. Note expanded Y-axis and compressed X-axis compared with Figures 4 and 5.**

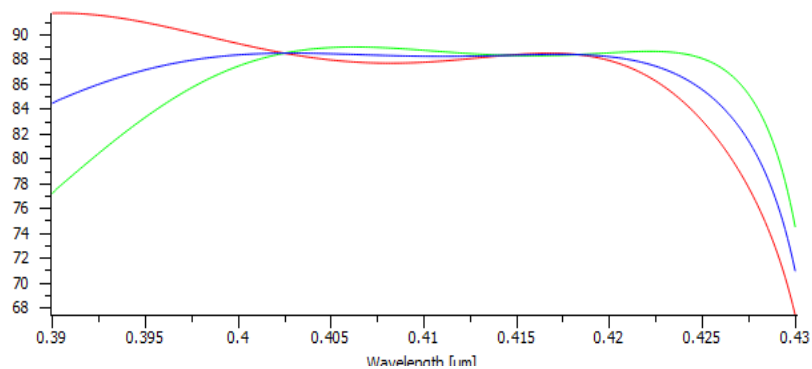

**Figure 8. Theoretical performance of a proposed 6450/mm planarized surface relief grating from Fraunhofer IOF. The efficiency for unpolarized light is shown in blue. Note expanded Y-axis and compressed X-axis compared with Figures 4 and 5.**

## 5. AN EFFICIENT AND COMPACT HIGH RESOLUTION DESIGN FOR MSE

A nominal design taking advantage of the gratings discussed in this paper has been laid out with 210mm beam. The design allows the use of either VPH or SR gratings. VPH gratings would be a (0,-1) grating as discussed in Section 3 for the Blue/Green arm, and a (1,-1) grating for the Red arm. SR gratings would have 85+% efficiency at all wavelengths in all arms, The slit has 8.7mm lateral smile to straighten the monochromatic slit image, and is gelled to a field lens. The collimator correctors are bonded to the grism input faces, saving two air/glass surface. Cameras are f/1.2 with 60mm x 60mm detectors. The largest lenses are 300mm aperture, the gratings are a plausible 500mm x 250mm. The overall physical size is 1.7m x 1.3m x 0.5m. The image quality is OK (Figure 10).

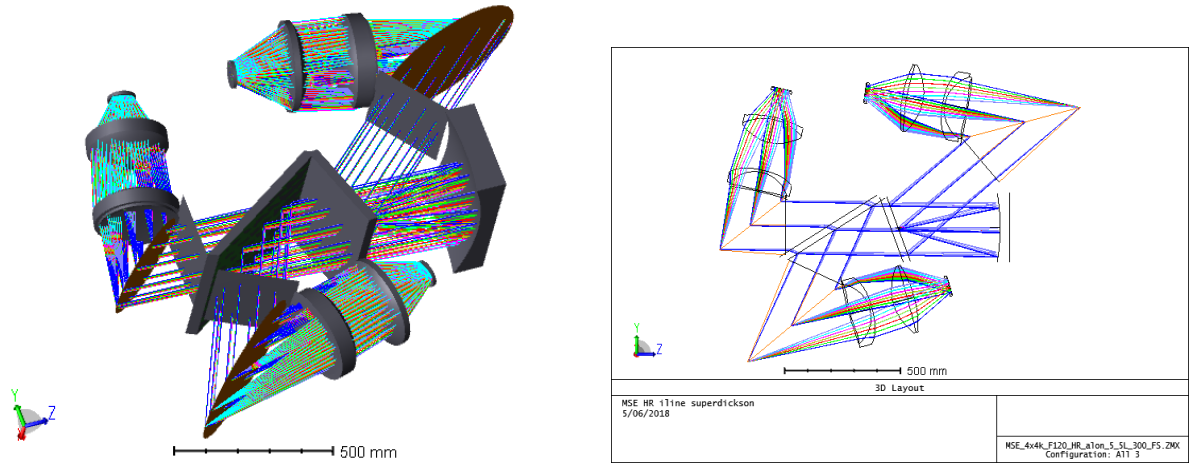

**Figure 9. Overall layout for the proposed design. Note the large prisms, for some reason not shown correctly on the ZEMAX solid model**

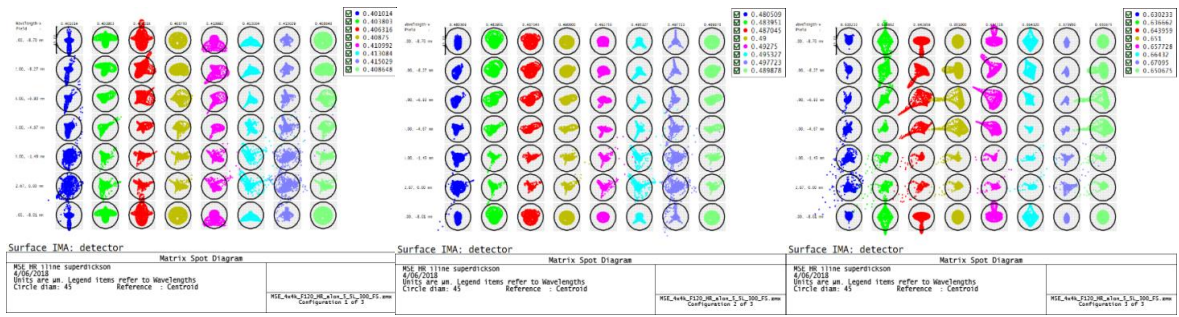

**Figure 10. Spot diagrams for the three arms. Circle size is 45μm, the projected fiber size on the detector.**

Only I-line glasses, fused silica and 4mm of AlON (the dewar window) are used, and the number of air-glass surfaces have been minimised. The principal losses are ~7% from overfilling the f/2.08 collimator, and ~7% from the slit obstruction. The overall efficiency (not including disperser) is 65-75% for each of the MSE HR spectral windows. With SR gratings, the MSE throughput requirements look feasible. Glass mass is about halved compared with the baseline design, and volume reduced by a factor of 3 or more.

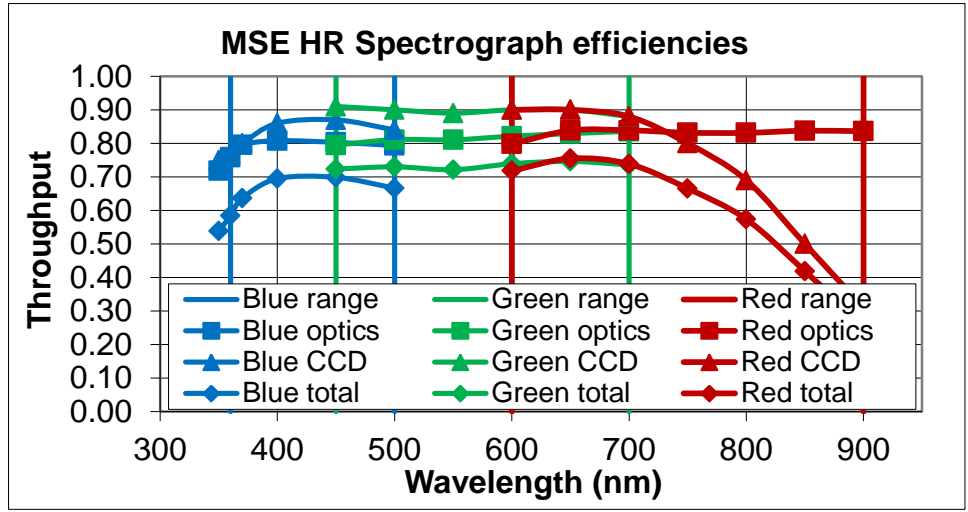

**Figure 11. Throughput for all spectrograph elements except disperser**

## 7. ACKNOWLEDGEMENTS

We have benefitted enormously from discussions and design effort by Jim Arns of KOSI, and we are extremely grateful for his candid help in developing solutions for this very challenging issue. We are also grateful to Andrea Zanutta for allowing a talk on these issues at short notice for the 'Dispersing Elements' workshop in Milan in October 2017, from which this paper arose. Thanks also to Aron Miller at SPIE for accepting the late manuscript.